\documentclass[journal=nalefd,manuscript=article]{achemso}

\usepackage{chemformula} 
\usepackage[T1]{fontenc} 
\usepackage{graphicx}
\usepackage{dcolumn}
\usepackage{bm}
\usepackage{mathrsfs}
\usepackage[T1]{fontenc}       
\usepackage[utf8]{inputenc}
\usepackage{lmodern}%
\usepackage{hyperref}%
\usepackage{xcolor}%
\usepackage{placeins}
\usepackage{tabularx}
\usepackage{booktabs}
\usepackage{mathtools}
\usepackage[normalem]{ulem}
\usepackage{wrapfig}
\usepackage{lipsum}

\newcommand{\ie}{\emph{i.e.},}

\newcommand{\mum}{\ensuremath{\mu}m}

\newcommand{\wse}{WSe\ensuremath{_2}}

\newcommand{\dqmp}{Department of Quantum Matter Physics, University of Geneva, 24 Quai Ernest Ansermet, CH-1211 Geneva, Switzerland}
\newcommand{\gap}{Group of Applied Physics, University of Geneva, 24 Quai Ernest Ansermet, CH-1211 Geneva, Switzerland}

\author{Daniil Domaretskiy}
\affiliation{\dqmp}
\alsoaffiliation{\gap}
\author{Nicolas Ubrig}
\affiliation{\dqmp}
\alsoaffiliation{\gap}
\author{Ignacio Gutiérrez-Lezama}
\affiliation{\dqmp}
\alsoaffiliation{\gap}
\author{Michael K. Tran}
\affiliation{\gap}
\author{Alberto F.Morpurgo}
\affiliation{\dqmp}
\alsoaffiliation{\gap}
\email{alberto.morpurgo@unige.ch}

\title[Identifying atomically thin crystals with diffusively reflected light]
  {Identifying atomically thin crystals with diffusively reflected light}

\definecolor{linkcol}{rgb}{0,0,0.4}
\definecolor{citecol}{rgb}{0.5,0,0}

\hypersetup{
	bookmarksopen=true,
	pdftitle="Identifying atomically thin crystals with diffusively reflected light",
	pdfauthor="Daniil Domaretskiy et al",
	pdfsubject="Identifying atomically thin crystals with diffusively reflected light", 
	pdfstartview={FitH},    
	pdfmenubar=true, 
	pdfhighlight=/O, 
	colorlinks=true, 
	pdfpagemode=UseNone, 
	pdfpagelayout=SinglePage, 
	pdffitwindow=true, 
	linkcolor=linkcol, 
	citecolor=linkcol, 
	urlcolor=blue
}

\begin{document}

 
\begin{abstract}
      The field of two-dimensional (2D) materials has been developing at an impressive pace, with atomically thin crystals of an increasing number of different compounds that have become available, together with techniques enabling their assembly into functional heterostructures. The strategy to detect these atomically thin crystals –based on optical contrast enhanced by Fabry-Pérot interference– has however remained unchanged since the discovery of graphene. Such an absence of evolution is starting to pose problems, because for many of the 2D materials of current interest the optical contrast provided by the commonly used detection procedure is insufficient to identify the presence of individual monolayers, or to determine unambiguously the thickness of atomically thin multilayers. Here we explore an alternative detection strategy, in which the enhancement of optical contrast originates from the use of optically inhomogeneous substrates, leading to diffusively reflected light. Owing to its peculiar polarization properties and to its angular distribution, diffusively reflected light allows a strong contrast enhancement to be achieved through the implementation of suitable illumination-detection schemes. We validate this conclusion by carrying out a detailed quantitative analysis of optical contrast, which fully reproduces our experimental observations on over 60 \wse\ mono-, bi-, and trilayers. We further validate the proposed strategy by extending our analysis to atomically thin phosphorene, $\mathrm{InSe}$, and graphene crystals. Our conclusion is that the use of diffusively reflected light to detect and identify atomically thin layers is an interesting  alternative to the common detection scheme based on Fabry-Pérot interference, because it enables atomically thin layers to be detected on substrates others than the commonly used Si/SiO$_2$, and it may offer higher sensitivity depending on the specific 2D material considered.
\end{abstract}


Following the discovery of graphene in 2005\cite{zhang_experimental_2005,novoselov_electric_2004}, the field of 2D materials has been continuing to develop extremely rapidly in many different directions\cite{geim_rise_2007,akinwande_two-dimensional_2014,novoselov_2d_2016}. Techniques have become available to manipulate atomically thin layers exfoliated from bulk crystals, which can now be stacked on top of each other with the desired orientation to form complex heterostructures\cite{dean_boron_2010,ponomarenko_tunable_2011,haigh_cross-sectional_2012,britnell_field-effect_2012,castellanos-gomez_deterministic_2014,cao_unconventional_2018}. The gamut of compounds produced in the form of atomically thin crystals has broadened enormously\cite{mounet_two-dimensional_2018,mak_atomically_2010,huang_layer-dependent_2017}, drastically expanding the scope of physical phenomena that can be accessed in these heterostructures\cite{novoselov_2d_2016,huang_lateral_2014,wang_tunneling_2018,rivera_observation_2015,ponomarev_semiconducting_2018,reddy_synthetic_2020,ubrig_design_2020,rizzo_charge-transfer_2020, wang_modulation_2020,yu_high-temperature_2019, zhao_sign-reversing_2019}. This impressive progress shows no sign of slowing down and a growing effort is devoted to employing systems based on atomically thin layers to realize structures of interest for future technological applications\cite{radisavljevic_single-layer_2011,koppens_photodetectors_2014,tao_silicene_2015,manzeli_2d_2017}.

What has enabled such an exceptionally fast progress in the field of 2D materials --and made the initial discovery of graphene possible-- is the use of optical microscopes to rapidly detect and locate crystals of many different compounds. The strategy relies on substrates covered with a sequence of layers acting as a Fabry-Pérot cavity –typically a Si wafer covered by a SiO$_2$ layer, whose thickness is optimized depending on the material to be detected --that enhance the contrast of atomically thin crystals in the visible spectrum\cite{blake_making_2007,jung_simple_2007,ni_graphene_2007,castellanos-gomez_optical_2010,benameur_visibility_2011}.  For a monolayer of graphene under optimized conditions, an intensity contrast as large as $\approx$~5\% can be reached, which makes the visualization of the layers straightforward. The thickness of multilayers can also be identified, because the difference in contrast produced by crystals whose thickness differ by an individual monolayer is larger than the \emph{noise} (\ie\ of random contrast variations measured on distinct layers of a same thickness). 

It seems remarkable that the impressive progress in the field of 2D materials has not led to any substantial change to the way in which the thickness of atomically thin crystals is determined based on optical contrast. This lack of evolution is starting to pose problems for multiple reasons. For many of the atomically thin layers that have been explored more recently, for instance, the optical contrast on commonly used Si/SiO$_2$ substrates does not provide sufficient sensitivity to determine the thickness unambiguously\cite{gorbachev_hunting_2011,philippi_lithium-ion_2018,alam_lithium-ion_2020,zhao_thickness_2020,couto_transport_2011}. Also, an increasing number of experiments rely on the use of specific substrate materials, but atomically thin crystals exfoliated on those substrates cannot be detected because their optical contrast is too small, with a notable exception being the use of optical transmission-mode to observe atomically thin crystals on transparent substrates routinely used in the assembly of heterostructures\cite{taghavi_thickness_2019}. For these reasons, it is important to explore new  ways to enhance the sensitivity for imaging and detecting atomically thin crystals of different materials under different experimental conditions. 

With this goal in mind, here we demonstrate a strategy to enhance the optical contrast of 2D materials, which relies on light that is diffusively reflected by substrates with optical inhomogeneous properties (\ie\ inhomogeneous refraction index). Exploiting the peculiar polarization properties of the diffusively reflected light, as well as its angular distribution, allows the contrast of atomically thin crystals to be optimized by employing different illumination and detection schemes. In particular, a large enhancement of contrast is found for a cross-polarization detection scheme, \ie\ letting the reflected light pass through polarizers oriented perpendicularly to the polarization of the incident light. We additionally reveal an unusually high sensitivity of the optical contrast to the numerical aperture (NA) of the microscope objective, which determines the amount of diffusely reflected light that is collected. We validate the proposed methods by investigating the optical contrast of more than 60 \wse\ mono-, bi-, and trilayers and we reproduce our experimental observations quantitatively by carefully modeling the details of diffusive reflection. We further show that all basic aspects of the technique  also  work for different 2D materials, such as black phosphorous, InSe, and graphene, indicating the  rather broad applicability of the proposed methods.  These conclusions are interesting, because the proposed strategy for the detection of atomically thin crystals is based on principles different from those employed in the existing procedure. It may therefore be applied to situations in which the existing procedure does not work, thereby contributing to broadening the scope of fabrication processes of structures based on 2D materials.

\begin{figure*}
		\centering
		\includegraphics[width=\linewidth]{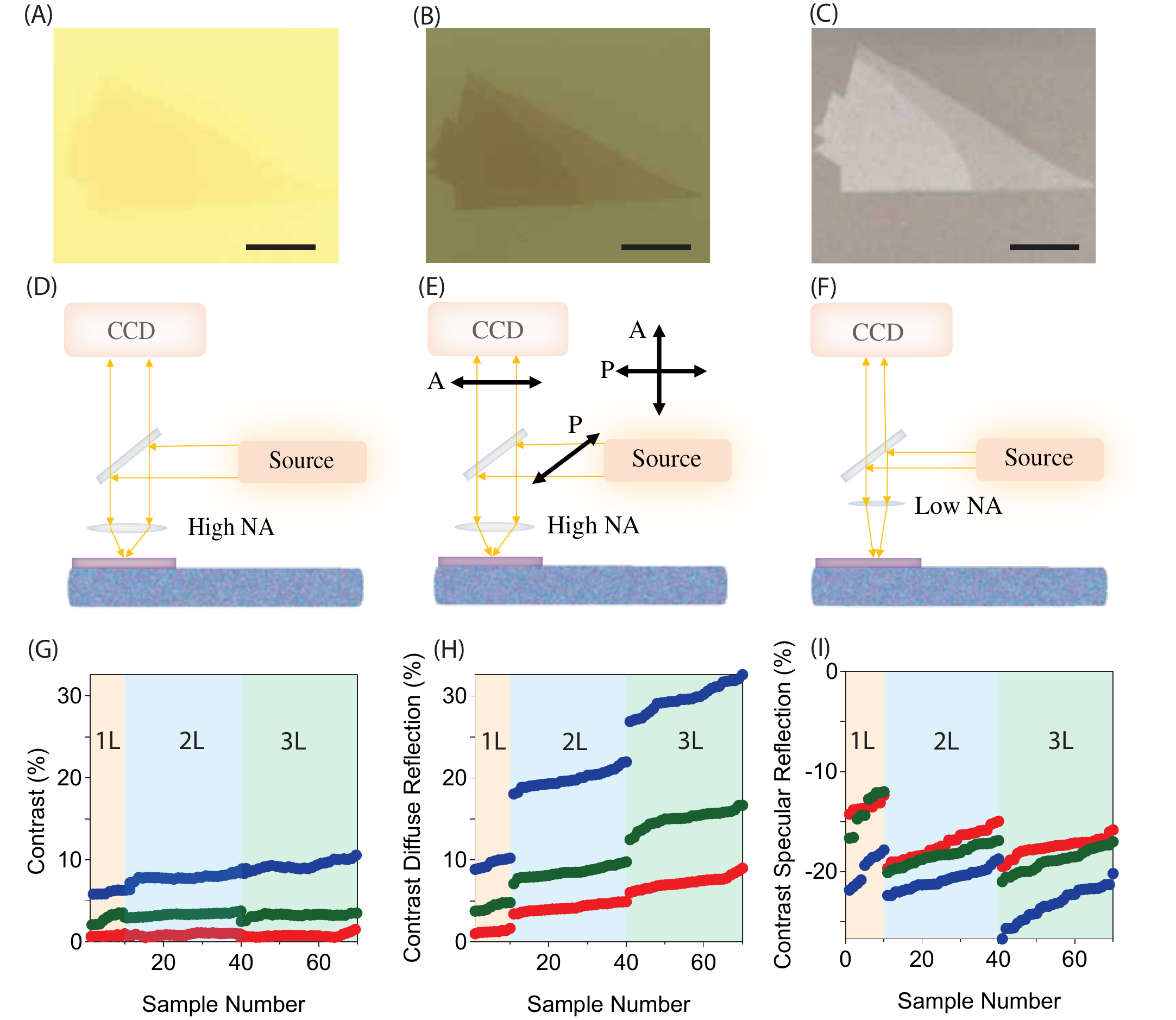}
		\caption{Optical contrast of \wse\ thin layers on glass-ceramic substrates under different illumination-detection conditions: image taken (A) under the \emph{conventional}\cite{li_rapid_2013,wang_thickness_2012} conditions illustrated in (D). (B) Image taken with the cross-polarization illumination-detection scheme illustrated in (E). Here the incident light is polarized with a polarizer P (s-pol), while the reflected light is analyzed with another perpendicularly oriented polarizer A. (C) Image captured with the illumination-detection with a low numerical aperture objective illustrated in (F). In all images, the scale bar corresponds to 20~\mum. The optical contrast of 1L, 2L, 3L \wse\ (more than 60 different crystals) measured in the three channels (Red, Green, and Blue) of the CCD is shown in (G) for conventional imaging conditions, (H) for the cross-polarization illumination-detection scheme, and in (I) for illumination-detection with a low numerical aperture objective.}
		\label{fig:01}
	\end{figure*}

Our work is motivated by the unexpected finding that the optical contrast of atomically thin \wse\ crystals on a glass-ceramic substrate very strongly depends on details of the imaging conditions, as illustrated in Fig.~\ref{fig:01}A-C. A glass-ceramics is an inhomogeneous material formed by a matrix of different oxides, causing its physical properties to be spatially inhomogeneous. What is important in the present context is that the microstructure of the glass ceramic that we use (LIC-GC, Ohara corparation)\cite{nakajima_lithium_2010} causes the refractive index to be inhomogeneous on a scale of 100-200~nm, i.e. a length scale comparable to the wavelength of light in the visible range. Fig.~\ref{fig:01}A shows the image of an atomically thin \wse\ crystal (containing and bi and a trilayer region) exfoliated on such a glass ceramic substrate, taken using an optical microscope with a high numerical aperture (NA~=~0.8) objective to both focus the light from a white source onto the substrate and to collect the reflected light. The reflected light is then recorded into an optical image using a three-channel (Red, Green, Blue) CCD-camera, according to the scheme illustrated in the Fig.~\ref{fig:01}D. This illumination-detection scheme is the one commonly employed to detect exfoliated crystals on Si/SiO$_2$ substrates, and we refer to it hereafter as \emph{conventional}\cite{li_rapid_2013,wang_thickness_2012}. Fig.~\ref{fig:01}A shows that the \wse\ layer is barely visible when imaged in this way, making it impossible to identify the regions of different thickness. 
	
If we polarize the incident light in the s-state, and let the reflected light pass through a polarizer oriented in the perpendicular direction before reaching the CCD camera (as illustrated in Fig.~\ref{fig:01}E), we find that the image contrast is very strongly enhanced (see Fig.~\ref{fig:01}B). Such an enhancement is unexpected, because for the same atomically thin crystal on a Si/SiO$_2$ substrate, letting the incident and reflected light pass through such a configuration of cross polarizers would extinguish the image completely. Instead, on glass ceramic substrates the thin crystals --which look darker than the substrate-- are clearly visible and so are the regions of different thickness. Finally, we note that a drastic change in contrast –with the crystal that becomes brighter than the substrate– is observed when we modify the conventional illumination-detection (\ie\ without polarizing the incident and reflected light), by simply changing the numerical aperture of the objective from NA=0.8 to NA=0.3 (Fig.~\ref{fig:01}F). 

It is clear from Figs. 1 A-C that the observed contrast of an atomically thin crystal on a glass ceramic substrate can be very strongly affected in the absence of any Fabry-Pérot interference, by acting on the polarization of the detected light or on its angular distribution (which is what changing the numerical aperture of the objective does). To quantify the differences between the different illumination-detection schemes we analyze the contrast of more than 60 exfoliated mono, bi, and trilayer (1L-3L) \wse. The result of this analysis is illustrated in Fig.~\ref{fig:01}G-I, where we have defined the contrast (in each of the R, G and B channels, and for each illumination-detection scheme employed) as the difference of the intensity measured on the substrate and that measured on the layer, divided by the intensity measured on the substrate. For conventional imaging conditions (Fig.~\ref{fig:01}G), the optical contrast is weak and does not vary significantly upon varying the \wse\ thickness. The cross-polarisation illumination-detection scheme results in a much higher optical contrast (Fig.~\ref{fig:01}H), exhibiting a clear discretization upon varying the thickness of \wse\ by one monolayer.  As already noted, the contrast changes its sign (Fig.~\ref{fig:01}I) when the low NA~=~0.3 objective is used.

\begin{figure*}
		\centering
		\includegraphics[width=\linewidth]{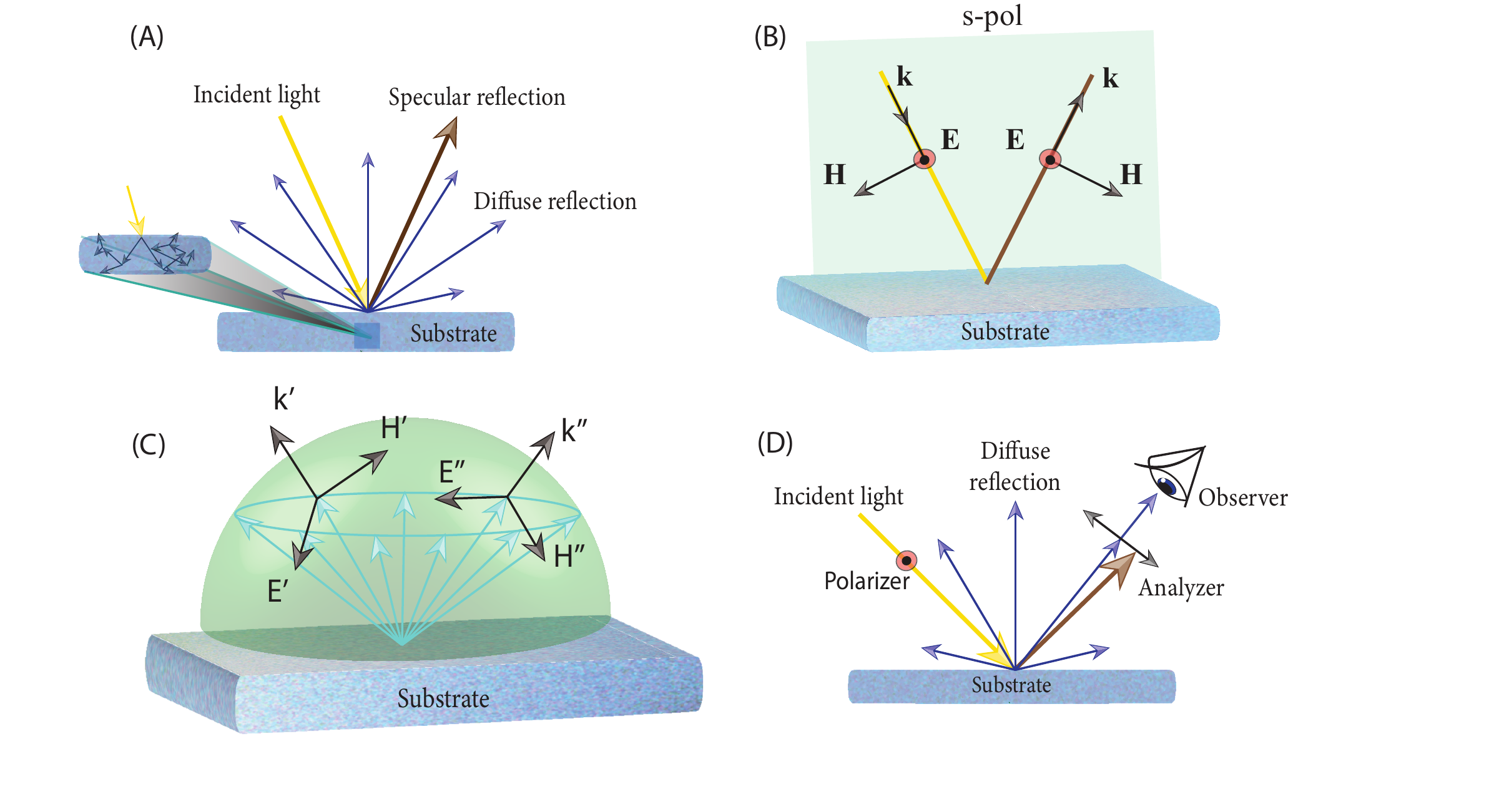}
		\caption{Schematic illustration of the different reflection channels and of their polarization properties, exploited in the different detection schemes. (A) conventional specular reflection (orange arrows) is accompanied by diffusive reflection (blue arrows) originating from the optical inhomogeneity of the substrate, which cause multiple internal reflection and refraction processes (as illustrated in the zoomed-in region). (B) Polarization of the incoming and reflected light in a conventional specular reflection process of an s-polarized beam. (C) Absence of polarisation of the diffusively reflected light, with \textbf{k$'$}, \textbf{k$''$} denoting the wave-vector of different partial waves of the diffusively reflected light, and \textbf{E$'$}/ \textbf{E$''$} and \textbf{H$'$}/ \textbf{H$''$} the corresponding electric and magnetic fields. (D) Specularly reflected light is completely extinguished by the analyzer in the cross-polarization illumination-detection scheme, which allows us to detect only the diffusively reflected non-polarized light from the substrate.}
		\label{fig:02}
\end{figure*}

The mechanism of this unusual sensitivity to the polarization (and to the numerical aperture of the objective used) originates from light that is diffusively reflected from the glass ceramic substrate, as we now first discuss qualitatively. For homogeneous substrates such as Si/SiO$_2$ incident light is reflected specularly, according to the usual Fresnel laws of reflection and refraction. On the contrary, the inhomogeneous microstructure of the glass ceramic substrate causes part of the incident light to undergo multiple reflection processes inside the substrate itself, before being eventually reflected with a wide angular distribution (Fig.~\ref{fig:02}A). This process –which is what we refer to as diffusive reflection– occurs in parallel to the conventional specular reflection determined by the average refractive index of the substrate. The intensity associated to the two channels –diffusive or specular reflection– is determined by the details of the microstructure in the material. The two reflection channels can be detected separately from each other by using their polarisation and angular distribution properties. Indeed, specularly reflected light is completely extinguished in the cross-polarisation scheme (Fig.~\ref{fig:01}E), because for incident light in the s-polarized state, specularly reflected light always preserves the polarization state (Fig.~\ref{fig:02}B). Diffusively reflected light, instead, has no polarisation due to multiple scattering processes on the microstructure (Fig.~\ref{fig:02}C), and therefore half of the intensity of the diffusively reflected light passes through the polarizer on the detection path (Fig.~\ref{fig:02}D). In addition, the use of low numerical aperture objective --which collects light from narrow angles-- results in a predominant detection of the specularly reflected light. Indeed, under normal incidence conditions, all specularly reflected light is detected with a low numerical aperture objective, whereas most of the diffusively reflected light --which has a much broader angular distribution-- is not. It is such a separation of two reflection channels that is responsible for the drastic differences observed in the images shown in Fig.~\ref{fig:01}A-C.

To confirm the scenario just outlined, we perform a quantitative analysis of diffusely reflected light in terms of its angular distribution and polarization properties, which enables us to calculate the intensity of the light detected by the CCD camera mounted on our microscope. Reflections and refractions of light in an inhomogeneous medium, whose refractive index changes randomly from its average value, can be described in terms of scattering cross section per-unit-volume $\sigma\left(\varphi\right)/V$, using an expression developed in the study of the physics of atmosphere (where electromagnetic radiation propagates through mist, clouds, turbulent flows of air, or under other conditions that locally change the refractive index)\cite{ishimaru_electromagnetic_2017}. This expression reads:
\begin{equation}
    \frac{\sigma\left(\varphi\right)}{V}=\frac{2}{\pi} \frac{1}{l}\ \frac{\left(kl\right)^4{\cos}^2\left[{\widetilde{n}}^2\varphi +\ \frac{\pi}{2}\left({\widetilde{n}}^2-1\right)\right]}{{\left[1+4\left(kl\right)^2{\sin}^2\frac{\varphi}{2}\right]}^2}
    \label{eqn:eq1}
\end{equation}
Here $\widetilde{n}$ quantifies the magnitude of the spatial inhomogeneity of the refractive index, $k$ is the wave-number of the incident light, $\varphi$ is the angle between the incident and scattered light, and $l$ is a correlation length of the refractive index in the inhomogeneous medium. 

To apply Eq.~(\ref{eqn:eq1}) to our glass ceramic substrates we take $l$ = 100~nm, corresponding to the characteristic size of the crystalline grains in the material. The magnitude of the variation in refractive index is $\widetilde{n}$~$\approx$~1, determined by the refractive indexes of the oxides that compose the material (Li$_2$O, Al$_2$O$_3$, SiO$_2$, P$_2$O$_5$, and TiO$_2$, see the material specification sheet\cite{nakajima_lithium_2010}) and by its microstructure, with TiO$_2$ particles having $n\approx2.7$ that are dispersed in a matrix of other oxides having $n\approx1.6$ (see also our own ellipsometry measurements discussed below). Eq.~(\ref{eqn:eq1}) then becomes:
\begin{equation}
    \frac{\sigma\left(\varphi\right)}{V}=\frac{2}{\pi} \frac{1}{l}\ \frac{\left(kl\right)^4{\cos}^2\varphi}{{\left[1+4\left(kl\right)^2{\sin}^2\frac{\varphi}{2}\right]}^2}
    \label{eqn:eq2}
\end{equation}
This relation describes the angular distribution of the scattered light, that is reflected back towards the microscope objective. Because the electric field of the diffusively reflected light is perpendicular to the direction of propagation, such light has no polarization. Using Eq.~(\ref{eqn:eq2}), we can directly compute the total intensity of the diffusive light that we collect with the NA=0.8 objective (a cone with the opening angle of $\mathrm{\sin^{-1}(NA)}$:

\begin{equation}
    I_D=\frac{2\pi}{{\frac{1}{2}4\pi r}^2}I_0\int_{\pi-\sin^{-1}{\left(0.8\right)}}^{\pi+\ \sin^{-1}{\left(0.8\right)}}V\frac{2}{\pi} \frac{1}{l}\ \frac{\left(kl\right)^4{\cos}^2\varphi}{{\left[1+4\left(kl\right)^2{\sin}^2\frac{\varphi}{2}\right]}^2}d\varphi
    \label{eqn:eq3}
\end{equation}
where $r$ = 5 mm is the focal distance of the objective, $I_0$ is the intensity of the incoming light. The factor 2$\pi$ at the numerator results from the integration over the polar angle, and the factor  $\frac{1}{2}$ accounts for the fact that only back-scattered light is collected. The scaterring volume V is determined as the product of the area of the field of view (a circle of radius 170~$\mu$m) and the substrate thickness (150~$\mu$m). Eq.~(\ref{eqn:eq3}) is the total intensity of diffusively reflected light that reaches the camera; if a polarizer is included in the detection path, as shown in Fig.~\ref{fig:01}E, the intensity is reduced by a factor of 2. Eq.~(\ref{eqn:eq3}) gives a quantitative prediction for the reflectance of the diffuse light $I_D/I_0$ as a function of wavelength, which can be tested experimentally. 

\begin{figure*}
		\centering
		\includegraphics[width=\linewidth]{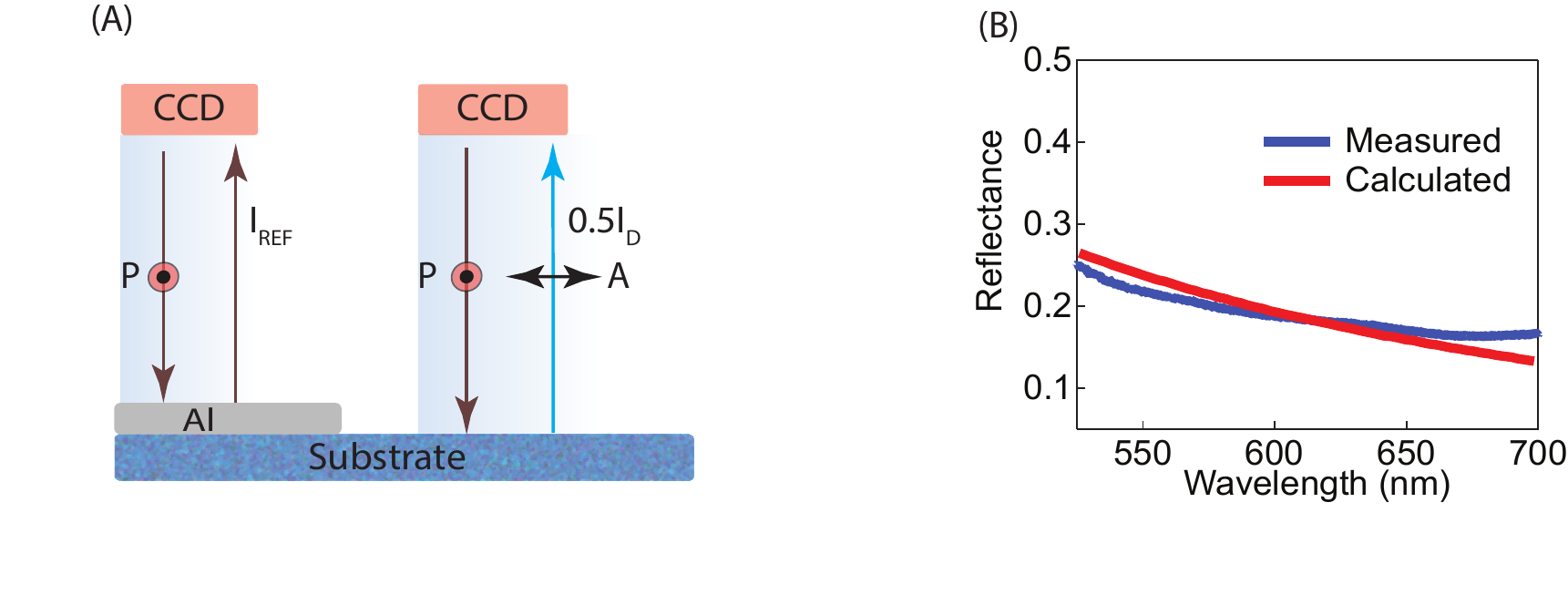}
		\caption{Measuring the reflectance of diffusively reflected light. (A) Schematics of the reflectance measurements, in which the intensity of the diffusively reflected light (right) is compared to the intensity of the light reflected from a 200~nm thick layer of Aluminum, which acts as a virtually perfect mirror. The incident light is polarized with a polarizer P (s-pol). The reflected light from the glass-ceramics is analyzed with another perpendicularly oriented polarizer A. (B) Experimentally determined (blue line) and theoretically calculated reflectance spectra of the diffusively reflected light.}
		\label{fig:03}
\end{figure*}

To this end, we perform reflectance measurements of diffuse light as a function of wavelength. We use a dedicated glass ceramic substrate that is covered in part with a sputtered 200~nm thick Al film, which acts as a virtually perfect mirror in the entire visible range.  Measuring the intensity of the light reflected from the Al layer allows $I_0$ to be measured, whereas the intensity reflected by the ceramic glass itself -- for the same intensity of the source -- gives us the value of I$_D$ (see the scheme in Fig.~\ref{fig:03}A). In practice, the Al reference is illuminated with an s-polarized beam of white light focused with a NA~=~0.8 objective. The reflected light is collected with the same objective and analyzed with the CCD-camera of a spectrometer. The same illumination procedure is used to illuminate the bare glass-ceramics, but the light diffusively reflected from the glass-ceramics is sent to the spectrometer after passing through a polarizing filter oriented perpendicularly to the polarization direction of the incident light (which fully eliminates the intensity due to the light originating from specular reflection). The ratio of $I_D$ and $I_0$ measured as a function of wavelength is shown in Fig.~\ref{fig:03}B (blue curve) and compared to the reflectance calculated using Eq.~(\ref{eqn:eq3}) (red curve).  The excellent quantitative agreement confirms the proposed scenario of diffusively reflected light and validates our quantitative analysis.

\begin{figure*}
		\centering
		\includegraphics[width=\linewidth]{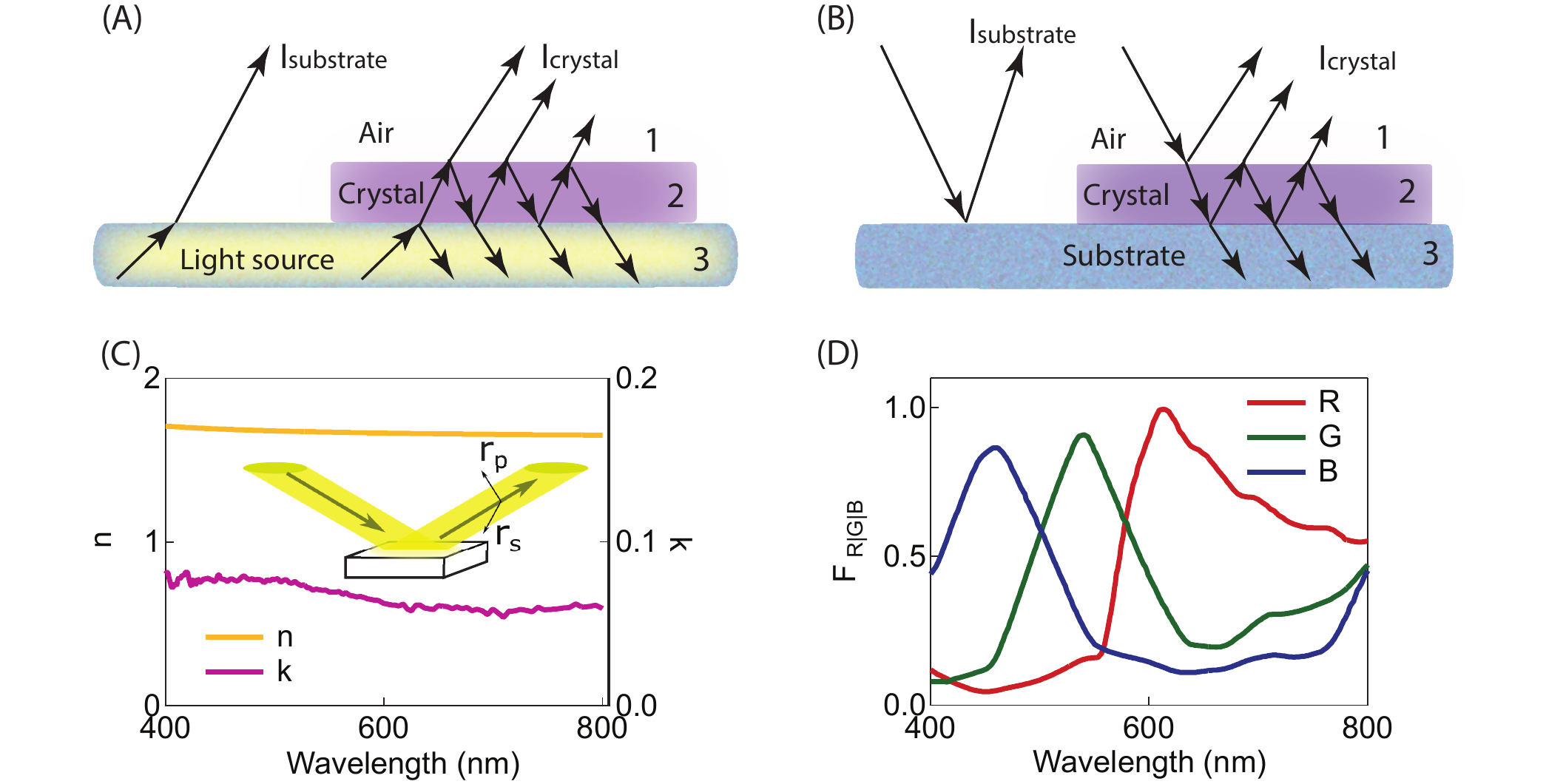}
		\caption{Calculation of the optical contrast for the diffusively and specularly reflected light. (A) Schematics of an atomically thin crystal (region 2) placed on a glass-ceramic substrate (region 3) in air (region 1). Calculations assume that the diffusively reflected light can be treated as a light source inside the substrate, emitting with the angular distribution given by Eq.~(\ref{eqn:eq2}), as discussed in the main text. The contrast can then be calculated as a transmission problem across the atomically thin crystal. (B) Multiple reflections at the substrate-crystal-air interfaces are considered to calculate the contrast of the specularly reflected part of the incident light. (C) Wavelength-dependent, complex refractive index ($n +ik$) of the glass-ceramics, as obtained from ellipsometry measurements. The inset illustrates the geometry of the measurements, where the ratio of $r_p$ and $r_s$ (Fresnel reflection coefficients for a linearly polarized incident light at the angle of incidence of 72$^{\circ}$) is used to obtain $n$ and $k$. (D) Spectral response functions of our CCD camera in the Red, Green, and Blue channels.}
		\label{fig:04}
\end{figure*}
Having established that diffuse light originates from the inhomogeneity of the refractive index, we proceed to compute the optical contrast of 2D crystals on a glass-ceramics substrate, to check if we can quantitatively reproduce the observations made in the different illumination-detection schemes (see Fig.~\ref{fig:01}). We describe the diffusively reflected light originating from the multiple reflection and refraction processes caused by the substrate inhomogeneities inside the substrate as a fictitious source of light (also positioned inside the substrate), which emits light with the angular distribution given by $\sigma\left(\varphi\right)/V$ (see Eq.~(\ref{eqn:eq2})). The contrast that we detect in the diffusive reflection channel (\ie\ what we measure when we insert an analyzer in the detection path) corresponds then to calculating the transmission of light emitted by this fictious source in the presence ($T_{321}$) and absence ($T_{31}$) of an atomically thin crystal at the surface of the substrate, and can be written as:

\begin{equation}
    C_{D}=\frac{T_{31} - T_{321}}{T_{31}}
    \label{eqn:eq4}
\end{equation}
The indexes 1, 2, and 3 indicate respectively air (complex refractive index - $n_1$), an atomically thin crystal (complex refractive index - $n_2$), and the substrate (complex refractive index - $n_3$; see Fig.~\ref{fig:04}A), so that ($T_{321}$) is the transmittance from the substrate to air passing through the atomically thin crystal and ($T_{31}$) is the transmittance from substrate to air. For normal incidence, the transmittance from the substrate to air $T_{31}=Re\left\{\frac{n_1}{n_3}\right\}{{|t}_{31}|}^2$, where $t_{31}=\frac{2n_3}{n_3+n_1}$ is Fresnel's transmission coefficient.  The transmittance of the substrate-thin crystal-air system reads\cite{stenzel_physics_2016,alma991010891849705251} (also for normal incidence): 
\begin{equation}
    \begin{aligned}
        T_{321}&=\frac{\frac{n_1}{n_3}{{\lvert}t_{32}\rvert}^2{{\lvert}t_{21}\rvert}^2\exp{\left(-\alpha d\right)}}{1 - 2\lvert r_{32}\rvert\lvert r_{21}\rvert \exp{\left(-\alpha d\right)} + {{\lvert r_{32}\rvert}\rvert}^2{\lvert r_{21}\rvert}^2\exp{\left(-2\alpha d\right)}}\\
        &\approx\frac{n_1}{n_3}{{\lvert}t_{32}\rvert}^2{{\lvert}t_{21}\rvert}^2\left[\frac{1}{{(1-\lvert r_{32}\rvert{\lvert}r_{21}\rvert)}^2}+ \frac{{(1+\lvert r_{32}\rvert{\lvert}r_{21}\rvert)}^3}{{(1-\lvert r_{32}\rvert{\lvert} r_{21}\rvert)}^2}{\alpha d}\right] \end{aligned}
        \label{eqn:eq5}
\end{equation}
(here $r_{ab}=\frac{n_a-n_b}{n_a{+n}_b}$ denotes Fresnel's reflection coefficient, $\alpha$ is the optical absorbance of the thin flake of thickness $d$, and we Taylor expand  $\mathrm{exp{\left(-\alpha d\right)}\approx 1-\alpha d}$ since $\mathrm{\alpha d\ll1}$ for atomically thin layers). We have checked that the Fresnel coefficients in Eq.~(\ref{eqn:eq5}) are nearly constant for the incident angles accessible with the NA~=~0.8 for unpolarized light\cite{alma991010891849705251}, so that the transmission process does not alter the angular distribution of the emitted light). Eq.~(\ref{eqn:eq5}) can then also be used to calculate quantitatively the optical contrast with the microscope objectives used in our experiments. 

The contrast that we measure in the specularly reflected channel (\ie\ when light is detected with a low numerical aperture objective to cut out most of the incidence angles, and without using a cross-polarization scheme) can be obtained using the usual relation for the contrast:
\begin{equation}
    C_{S}=\frac{R_{13} - R_{123}}{R_{13}}
    \label{eqn:eq6}
\end{equation}
where $R$ stands for reflectance and the subscripts have the same meaning as described above for the transmittance. Indeed, specular reflectance of the substrate-air and the substrate-thin crystal-air system (Figs.~\ref{fig:04}B) is a well-known problem that we model with the Fresnel equations for a normal incidence\cite{stenzel_physics_2016,alma991010891849705251}, which depends only on the average refractive index (the inhomogeneity of the substrate plays no role). We obtain:

\begin{equation}
    \begin{split}
        R_{123}&=\frac{{\lvert r_{12}\rvert}^2-2\lvert r_{12}\rvert\lvert r_{23}\rvert \exp{\left(-\alpha d\right)}+{\lvert r_{23}\rvert}^2\exp{\left(-2\alpha d\right)}}{1 - 2\lvert r_{12}\rvert\lvert r_{23}\rvert \exp{\left(-\alpha d\right)} + {\lvert r_{12}\rvert}^2{{\lvert r}_{23}\rvert}^2\exp{\left(-2\alpha d\right)}}\\
        &\approx\left[\frac{{(\lvert r_{12}\lvert\ - \ \lvert r_{23}\lvert)}^2}{{(1-\lvert r_{12}\rvert{\lvert r}_{23}\rvert)}^2}-\frac{\lvert r_{23}\lvert(\lvert r_{12}\lvert\ - \ \lvert r_{23}\rvert)({\lvert r_{12}\rvert}^2+1)}{{(1-\lvert r_{12}\rvert\lvert r_{23}\rvert)}^3}{{\alpha d}}\right]
    \end{split}
    \label{eqn:eq7}
\end{equation}
from which $R_{13}$ can be obtained by setting $d$~=~0 and substituting the index $2$ with $1$ in the Eq.~(\ref{eqn:eq7}). We have implicitly assumed that the Fresnel reflection coefficient in Eq.~(\ref{eqn:eq7}) are the same as those that one would have for a uniform substrate with a refractive index equal to the average value of our (inhomogeneous) substrate. Clearly, this assumption cannot be exact because the diffusively reflected light takes part of the incident power. However, the excellent agreement that we find with experimental data (see below) shows that our assumption about the Fresnel coefficient corresponds to a good description of the experimental situation. This is probably because the total amount of intensity of the light reflected diffusively is relatively small, so the modification to the Fresnel coefficients is also small. 

To determine quantitatively the contrast, we need to know the transmission and reflection coefficients in the Equations \ref{eqn:eq5} and \ref{eqn:eq7}, which depend on the complex refractive indexes of the glass-ceramic substrate and of the \wse\ thin crystals. We obtain the complex refractive index of the substrate by performing wavelength dependent ellipsometry measurements (the geometry of the experiment is presented in the inset Fig.~\ref{fig:04}C). In these measurements the light spot has a size of approximately 3~mm and probes the average behavior of the refractive index; the diffusive part of the reflected light can be neglected due to a large distance between the substrate and the detector ($\approx$~30~cm; due to its broad angular distribution, only a negligible intensity due to diffusively reflected light is detected for such a large distance). The obtained refractive index of the class-ceramics is almost constant (n~$\approx$~1.65) for wavelengths in the visible range, with almost no absorption (Fig.~\ref{fig:04}C). The refractive indexes of \wse\ are taken from literature\cite{hsu_thickness-dependent_2019}. As a final step to calculate the measured contrast for the diffusive and the specular reflection channels we need to convolute the wavelength dependent quantities in Eqs.~\ref{eqn:eq4} and \ref{eqn:eq6} with the spectral response $F_R\left(\lambda\right)$, $F_G\left(\lambda\right)$, $F_B\left(\lambda\right)$ of the individual R, G, and B channels of a CCD camera that we use (given by the manufacturer and reproduced in Fig.~\ref{fig:04}C). We obtain: 
\begin{equation}
    C_{D,R\lvert G\rvert B}=\frac{\int{(T}_{31}\left(\lambda\right)\ - T_{321}\left(\lambda\right))F_{R\lvert G\rvert B}\left(\lambda\right)d\lambda}{\int{T_{31}\left(\lambda\right)F_{R\lvert G\rvert B}\left(\lambda\right)d\lambda}}
    \label{eqn:eq8}
\end{equation}
\begin{equation}
    C_{S,R\lvert G\rvert B}=\frac{\int{(R}_{12}\left(\lambda\right)\ - R_{123}\left(\lambda\right))F_{R\lvert G\rvert B}\left(\lambda\right)d\lambda}{\int{R_{12}\left(\lambda\right)F_{R\lvert G\rvert B}\left(\lambda\right)d\lambda}}
    \label{eqn:eq9}
\end{equation}
Eqs. (8) and (9) can then be directly compared to the experimental values, as we proceed to do next.

\begin{figure*}
		\centering
		\includegraphics[width=\linewidth]{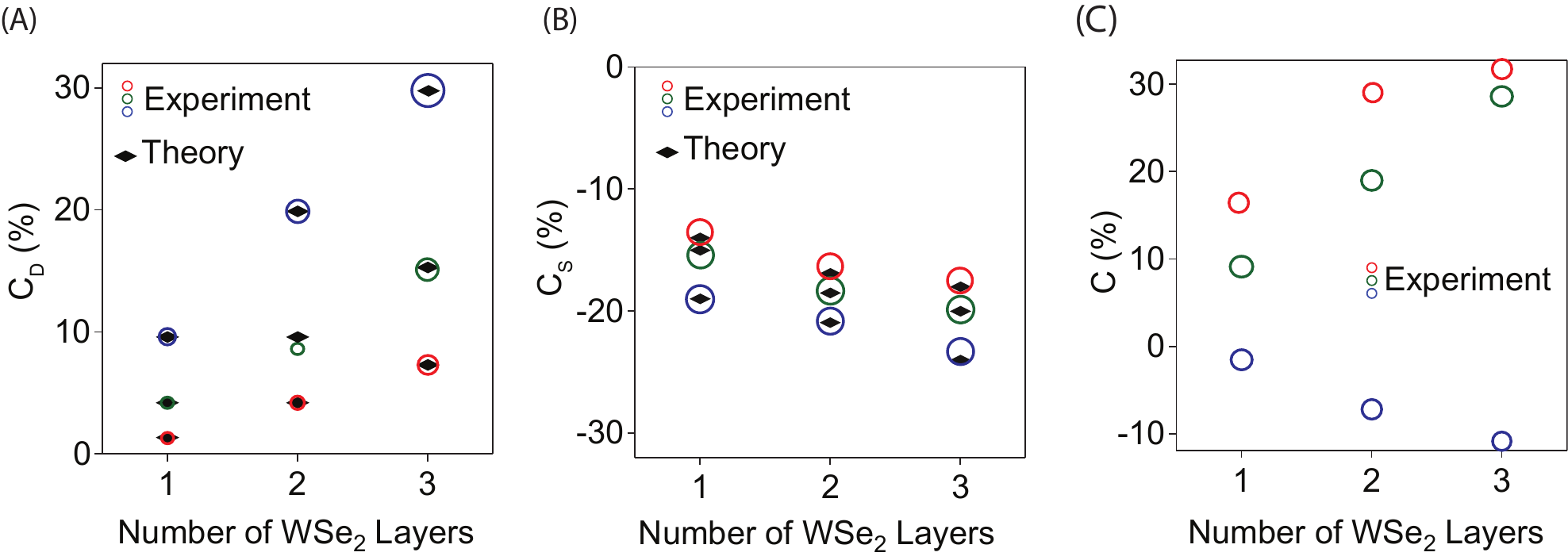}
		\caption{Quantitative analysis of the optical contrast of \wse\ layers in the different illumination/detection schemes. (A) Optical contrast of diffusively reflected light for mono-, bi-, and trilayers on glass ceramic substrates. The experimentally determined values of the contrast for the RGB channels are represented with red, green, and blue empty circles, whose size corresponds to the standard deviation of the contrast measured on many different samples. The theoretical values of the contrast are shown as black full symbols. (B) Optical contrast in the case of specularly reflected light symbols have the same meaning explained for (A). (C) Optical contrast of \wse\ crystals on Si/SiO$_2$ substrates with a SiO$_2$ thickness of 275~nm, optimal for detecting a few-layer \wse. }
		\label{fig:05}
\end{figure*}

We use the data from the Fig.~\ref{fig:01}E-F to calculate the average values of the contrast in the R, G, and B channels of the CCD camera for 1L-3L \wse\ crystals imaged under the different illumination-detection schemes. The experimental data obtained in this way are shown together with the theoretically calculated values in Fig.~\ref{fig:05}A and B, from which the excellent agreement between calculated and measured quantities is apparent. The contrast changes nearly linearly as predicted by Eq.~(\ref{eqn:eq5}) and (\ref{eqn:eq7}) (with deviations due to the strong dependence of the complex refractive index of \wse\ on the thickness of the crystal and the wavelength of incident light). The sensitivity that we obtain from the contrast measured in the diffusively reflected light channel is as large as –or possibly even better than– the contrast measured for the same crystals on Si/SiO$_2$ substrates with the conventional illumination-detection scheme, based on Fabry-Pérot interference (see Fig.~\ref{fig:05}C). Indeed, in the diffusively reflected light channel, the contrast allows to discriminate between 1L, 2L, and 3L in each of the individual R, G, and B channels. Overall, the quantitative agreement between the experimental values and the calculated ones fully confirms the relevance of diffusively reflected light and the way in which we have modelled the diffusive reflection process.

For completeness, we also analyze the optical contrast in the case in which light is detected from both diffusive and specular reflection channels, corresponding to imaging in the conventional detection scheme (see Fig.~\ref{fig:01}A and D). The total intensity of the light reflected from the substrate $I_{\rm sub}$  is the sum of intensity of the specularly and diffusively reflected light, respectively $I_{\rm sub,S}$ and $I_{\rm sub,D}$  (and the same is true for the light reflected from a thin crystal on the substrate, $I_{crys}=I_{crys,D} + I_{cryst,S}$). From the definition of optical contrast, we have:
\begin{equation}
    C =\frac{I_{\rm sub}-I_{crys}}{I_{\rm sub}}=\frac{I_{\rm sub,D}\ }{I_{\rm sub,D}+I_{\rm sub,S}}C_{D}+\frac{I_{\rm sub,S}\ }{I_{\rm sub,D}+I_{\rm sub,S}}C_{S}  
\label{eqn:eq10}
\end{equation}
$I_{\rm sub,S}$ is obtained from Fresnel law using the measured refractive index of the glass ceramic substrate and $I_{\rm sub,D}$ is known from the reflectance measurements discussed earlier (see Fig.~\ref{fig:03}). As Fig.~\ref{fig:01}A and 1G show that the optical contrast is nearly thickness independent, it is sufficient to analyze its value for the case of bilayer \wse\ . Using the values of $C_D$ and $C_S$ calculated theoretically (see Fig.~\ref{fig:05}) we find from Eq.~(\ref{eqn:eq5}) that in the Red channel C=0.2\%, in the Green channel C=2.0\%, and in the Blue channel C=9.6\%, in very good agreement with the experimental data shown in Fig.~\ref{fig:01}G. The values observed in this conventional imaging mode are the result of the competition between two reflection channels (Eq.~(\ref{eqn:eq10})), which contribute with opposite sign, thereby decreasing the total contrast and explaining the faintness of the crystal as imaged in Fig.~\ref{fig:01}A.

\begin{figure*}
		\centering
		\includegraphics[width=\linewidth]{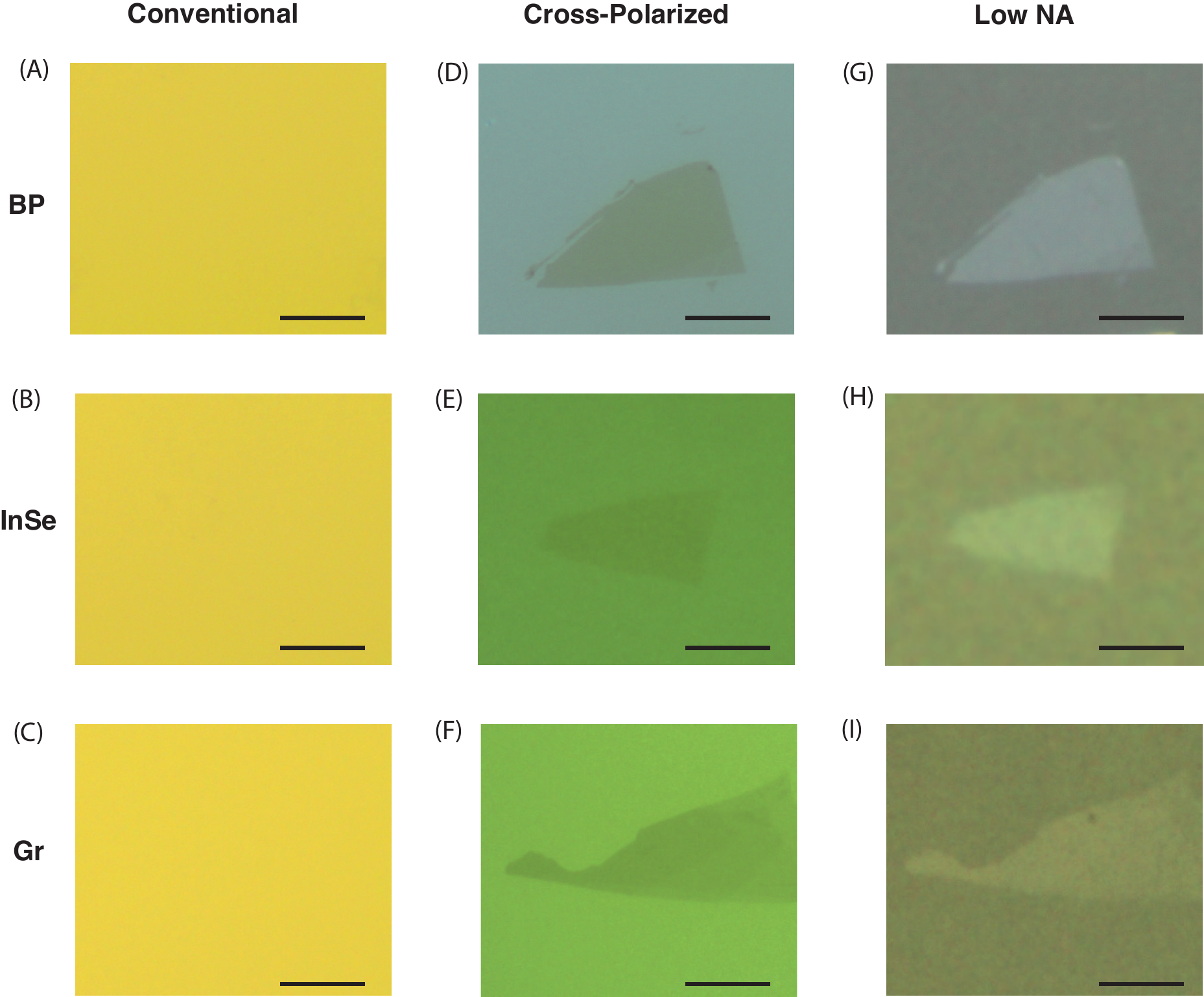}
		\caption{Optical contrast of black phosphorus (BP), InSe, and graphene thin layers on glass-ceramic substrates under different illumination-detection conditions. Panels (A)~-~(C) show images of the different materials taken under conventional illumination and detection conditions,  resulting in a virtually vanishing optical contrast. Images in panels (D)~-~(F) are taken using the cross-polarisation illumination-detection scheme. The atomically thin layers are darker than the substrate (positive contrast) and are  easily detected. Panels (G)~-~(I) show images taken  with a low numerical aperture objective, resulting in a negative contrast that also easily allows the detection of the atomically thin crystals. The overall behaviour of black phosphorous, InSe and graphene on a glass-ceramic substrate is therefore identical to that of WSe$_2$, and can be understood as due to  diffusively reflected light from the  optically non-uniform glass-ceramic substrate.   In all images, the scale bar corresponds to 10~\mum.
		}
		\label{fig:06}
\end{figure*}

Having established the role of diffusively reflected light and understood the origin of the contrast differences in the different imagining modes that we introduced  at the beginning (see Fig. 1D-F), we test whether the conclusions obtained for WSe$_2$ also hold true for other 2D materials. To this end, we look at  different families of materials (black phosphorus, $\mathrm{InSe}$, and graphene), and analyze images taken in the conventional imaging mode, with cross polarizers, and with low numerical aperture (see Fig. 6) to  check if all the key aspects of these different imaging modes are present irrespective of the material considered. We do not repeat the complete quantitative analysis and only focus our attention on layers of the same thickness --4L-- for the three materials. Conventional imaging of the crystals on a glass-ceramics  results in a weak contrast that does not allow spotting any crystals easily (Fig~\ref{fig:06}A-C) as in the case of \wse  (Fig~\ref{fig:01} A). Using the cross-polarisation illumination-detection scheme, instead, makes all crystals become clearly visible (Fig~\ref{fig:06}D-F). With this scheme, the crystals look darker than the substrate, just as WSe$_2$. The resulting  positive contrast is approximately 5$\%$ per layer for phosphorene and graphene, and 2$\%$ per layer for $\mathrm{InSe}$ in the Blue channel. Again similarly to the case of \wse, the crystals look brighter than the substrate when imaged under the illumination-detection scheme with low NA (Fig~\ref{fig:06}G-I). The negative contrast then is as high as 6$\%$ and 5$\%$ per layer for $\mathrm{InSe}$ and phosphorene, and 3$\%$ per layer for graphene (again, in the Blue channel). Therefore, the technique works for all materials tested and the sensitivity in optical contrast that can be obtained by exploiting diffusively reflected light  is at least as large as that given by Fabry-Perot interference  on Si/SiO$_2$ substrates in the conventional illumination-detection scheme. We therefore conclude that the use of diffusively reflected light in combination with one of the illumination-detection schemes discussed here is a promising alternative to identify many different exfoliated 2D crystals on a substrate, and determine their thickness.

\begin{acknowledgement}
The authors thank D.~van der Marel for fruitful discussions and A.~Ferreira for technical support. A.F.M. acknowledges financial support from the Swiss National Science Foundation (Division II) and from the EU Graphene Flagship project. 
\end{acknowledgement}

\section{Metod}

Thin layers of 2H-\wse, black phosphorus, $\gamma$-InSe, and graphene are mechanically exfoliated onto Si/SiO$_2$ substrates from bulk crystals of the same compounds (purchased from HQGraphene). The exfoliated crystals are then identified and imaged with an Olympus BX51M reflection-mode microscope under conventional illumination-detection conditions using a high NA Olympus 50x MPlanFL N objective.  The exfoliated layers are subsequently transferred onto glass ceramic substrates using an established all-dry transfer technique\cite{castellanos-gomez_deterministic_2014}. For air-sensitive materials such as InSe and black phosphorous the entire process is carried out in a nitrogen filled glove-box to avoid  degradation. Once placed onto glass ceramic substrates, the layers are imaged under conventional conditions, cross-polarization illumination-detection conditions (we use Olympus U-PO3 as a polarizer and Olympus AN360-3 as an analyzer), and with a low NA objective illumination-detection technique (Olympus 50x SLMPlan N objective). We then use a raster graphics editor (Adobe inc.) to extract intensities in the RGB channels of the images for the thin crystals and the substrates under all illumination-detection conditions. The extracted intensities allow us to compute optical contrast as described in the main text.

Reflectance specta are obtained by measuring the reflected light of a broadband Tungsten-Halogen light source focused onto the substrate, and by normalizing it to light reflected by a 200 nm Al layer sputtered onto part of the substrate, acting as  reference mirror (200 nm Al, see Fig. 3A and the main text for the exact description of the measurements). The light is sent to a Czerny-Turner monochromator (Andor Shamrock) and detected with a cooled Si-CCD array (Andor EmCCD).

\bibliography{contrast}

\end{document}